\documentclass[fleqn,usenatbib]{mnras}

\usepackage{newtxtext,newtxmath}

\usepackage[T1]{fontenc}

\DeclareRobustCommand{\VAN}[3]{#2}
\let\VANthebibliography\thebibliography
\def\thebibliography{\DeclareRobustCommand{\VAN}[3]{##3}\VANthebibliography}

\usepackage{graphicx}	
\usepackage{amsmath}	
\usepackage[]{threeparttable}
\graphicspath{{./}{figures/}}

\newcommand{\hii}{H\thinspace{\sc ii}}

\newcommand{\oiii}{O\thinspace{\sc iii}}
\newcommand{\ovi}{O\thinspace{\sc vi}}

\newcommand{\ciii}{C\thinspace{\sc iii}}
\newcommand{\civ}{C\thinspace{\sc iv}}

\newcommand{\nv}{N\thinspace{\sc v}}
\newcommand{\heii}{He\thinspace{\sc ii}}

\newcommand{\lya}{Ly$\alpha$ }

\title[Type~II QSO S82-20]{A Peculiar Type~II QSO Identified via Broad-band Detection of Extreme Nebular Line Emission
{\footnote{The LBT is an international collaboration among institutions in the United States, Italy and Germany.  LBT Corporation partners are: The University of Arizona on behalf of the Arizona university system; Istituto Nazionale di Astrofisica, Italy; LBT Beteiligungsgesellschaft, Germany, representing the Max-Planck Society, the Astrophysical Institute Potsdam, and Heidelberg University; The Ohio State University, and The Research Corporation, on behalf of The University of Notre Dame, University of Minnesota and University of Virginia.}}
{\footnote{based on observations made with the Southern African Large Telescope (SALT)}}
}

\author[Y.  Lin et al.]{
Yu-Heng Lin$^{1, 2}$\thanks{E-mail: lin00025@umn.edu},
Claudia Scarlata$^{2}$,
Matthew Hayes$^{3}$,
Anna Feltre$^{4}$,
Stephane Charlot$^{5}$,
Angela Bongiorno$^{6}$,
\newauthor
Petri Väisänen$^{7, 8}$,
and Moses Mogotsi$^{7, 8}$
\\
$^{1}$School of Physics and Astronomy, University of Minnesota, 116 Church St SE, Minneapolis, MN 55455, USA\\
$^{2}$Minnesota Institute for Astrophysics, University of Minnesota, 116 Church St SE, Minneapolis, MN 55455, USA\\
$^{3}$Stockholm University, Department of Astronomy and Oskar Klein Centre for Cosmoparticle Physics, AlbaNova University Centre, SE-10691, Stockholm, Sweden\\
$^{4}$INAF - Osservatorio di Astrofisica e Scienza dello Spazio di Bologna, Via P.  Gobetti 93/3, 40129 Bologna, Italy\\
$^{5}$Sorbonne Universit\'e, CNRS, UMR7095, Institut d'Astrophysique de Paris, F-75014, Paris, France\\
$^{6}$INAF - Osservatorio Astronomico di Rioma, Via Frascati 33 00074, Monteporzio Catone, Italy\\
$^{7}$South African Astronomical Observatory, P.O.  Box 9, Observatory, 7935, Cape Town, South Africa\\
$^{8}$Southern African Large Telescope, P.O.  Box 9, Observatory, 7935, Cape Town, South Africa\\
}

\date{Accepted 2021 October 11. Received 2021 September 27; in original form 2021 May 5}

\pubyear{2021}

\begin{document}
\label{firstpage}
\pagerange{\pageref{firstpage}--\pageref{lastpage}}
\maketitle

\begin{abstract}
We present S82-20, an unusual redshift $\approx$3 object identified  in SDSS-Stripe 82 broad-band images.  The rest-frame ultraviolet spectrum of S82-20 shows emission lines from highly ionized species, including \heii\ $\lambda$1640, and the \civ\ $\lambda\lambda$1548, 1550 and \ovi\ $\lambda\lambda$1032, 1038 doublets.  The high \lya luminosity ($3.5\times 10^{44}$ erg s$^{-1}$), the high emission line equivalent widths ($>200$\AA\ for \lya), the FWHM of the emission lines ($<800$km s$^{-1}$), and the high ionization \ovi\ line strongly support the interpretation that S82-20 is a Type~II QSO.  However, photoionization models using Type~II QSO do not fully explain  the measured \civ/\heii\ line ratio, which requires either some contribution from star-formation or high velocity shocks.  Additionally, S82-20 is not detected at wavelengths longer than 2$\mu$m, in tension with the expectation of isotropically IR emission of a luminous QSO. We consider the possibility that S82-20 is a rare example of a changing-look QSO, observed in a temporarily low state, where the broad line region has faded, while the narrow line region still emits emission line.  Otherwise, it may be a rare case of the short phase of the life of a massive galaxy, in which active star formation and accretion onto a supermassive black hole coexist.
\end{abstract}




\section{Introduction}

Observations suggest that the cosmic evolution of the star formation rate and black hole accretion rates are similar: both increase with redshift, reaching a peak at redshift $z \simeq 2$, and declining in the earlier universe \citep[e.g.,][]{Silverman2008, Madau2014, Vito2018}.  
This similarity, together with the fact that supermassive black holes (SMBHs) are ubiquitous in the centers of galaxies in the local universe as well as the well known scaling relations between galaxies' properties and black hole mass \citep{Magorrian1998} suggest the existence of a tight connection between the growth of SMBHs and the evolution of their host galaxy \citep{Ferrarese2000,Gebhardt2000,Madau2014}.  
It has been suggested that this connection results from the fact that the BH accreting phase contributes to the regulation of galaxy growth, possibly via feedback.  On one hand, BH activity traces gas accretion; on the other hand, this phase is also a source of feedback (for example, through powerful winds) and may be able to quench star-formation in massive haloes \citep{Hopkins2008, Richardson2016}.

SMBHs can be identified and studied while they are accreting material.  This phase is called the Active Galactic Nucleus (AGN) phase.  In the unification model, the accretion disk is surrounded by an obscuring torus \citep{Antonucci1993, Urry1995}.  This simple model explains a vast range of observations, as the result of viewing conditions.  In this context, AGN are divided into Type~I and Type~II AGN, depending on whether or not the broad line region (which is generated in the centermost regions) is visible.  For Type~II, obscured AGN, the line of sight to the broad-line region is blocked by the dusty torus, while the narrow-line region, which can extend up to scales of several kpcs, can still be observed.  

In addition to the unification model, the obscured phase can be explained with the merger-driven scenario \citep{DiMatteo2005, Hopkins2006, Hopkins2008}.
In this scenario, gas-rich galaxy mergers drive inflows, producing central starbursts and fueling the growth of the SMBH.  Type~I and Type~II AGN are then objects observed in different evolutionary phases rather than from different angles .  During the obscured phase, the nucleus is buried in the surrounding inflowing materials.  After the energy released in the accretion process dominates in the central regions, radiative feedback can blowout the surroundings and the active galaxy evolves into an unobscured quasar.  
Some AGNs were observed to dramatically change their broad emission component, with the spectra varying between Type I, Type II, or the intermediate Type 1.5, 1.8 \citep{Penston1984, VO1987, Elitzur2014}.  The variation of these so called “changing-look” AGN can be driven by different mechanisms including: (1) variable obscuration due to a moving patchy dust torus \citep{Elitzur2012}; (2) variable accretion rate in the central engine \citep{Elitzur2014, LaMassa2015}; (3) tidal disruption events \citep[e.g.,][]{Eracleous1995}.

High redshift Type~I quasars are routinely identified in optical surveys thanks to their colors \citep{Padovani2017}.  Spectroscopic confirmation is then easily achieved, given their brightness, and easy-to-detect broad emission lines in the rest-frame UV.  
Optical surveys, however, are less successful at identifying Type~II QSOs, since the AGN continuum, as well as the broad emission lines, are completely obscured along our line of sight.  Therefore, the identification of Type~II QSOs is a multi-step process that requires spectroscopic follow-up to determine the nature of the source responsible for the excitation of the gas.  

Many surveys have searched for Type~II QSOs, using criteria based on the interaction between photons and the obscuring material.  
Typically, Type~II obscured accretion is identified in the following ways: X-ray observations detect the obscured high-energy photons, mid- to far-IR observations directly probe the thermal re-emission of dust, and optical spectroscopic observations search for narrow emission lines with no broad components \citep[e.g.,][]{Lacy2004, Donley2012, Stern2012, Assef2013, Hainline2014, Alexandroff2013, Hickox2017}. 
Recently, however, it has become clear that obscured accretion may look quite different, depending not only on the column density of obscuring material, but also on the intrinsic luminosity \citep[e.g.,][]{Ueda2003,Hasinger2004} and the BH accretion rate \citep[e.g.,][]{Fabian1999}.  Additionally, different mechanisms may originate obscured activity \citep{Rigby2006, Li2020}.  Thus, alternative methods to identify obscured quasars are important to truly constrain the variety of the AGNs and their correlations to the host galaxies.

Spectroscopic searches for high redshift quasars use the presence of bright and narrow emission lines in the rest-frame UV to identify Type~II quasar candidates \citep[e.g., ][who applied a cut at line widths Full Width Half Maximum of 2000 km s$^{-1}$]{Alexandroff2013}.  These samples result in moderately obscured quasars, as revealed by follow-up rest-frame optical spectroscopy \citep{Greene2014,Hickox2018, Burtscher2016, Schnorr-Muller2016}.

A promising complementary technique to identify Type~II QSOs is to search for objects with bright UV emission lines but a faint continuum.  This is because, in the unification model, the bright continuum from the accretion disk and the broad-line region would be fully obscured by the torus along the line of sight, while the kpc-scale narrow-line region would still shine unaffected \citep[e.g., ][]{Antonucci1993,Urry1995,Nakajima2018}.  
Narrow, high EW emission lines can however also be a feature of young metal-poor star-forming galaxies (SFGs).  High EWs of the \lya\ emission line (e.g., larger than few hundreds \AA) are also expected for metal-free Population~III stars \citep{Schaerer2003, Raiter2010}.  

In a search for extreme emission line galaxies, we have identified a promising Type~II QSO candidate.
We are conducting a broad-band search for emission line galaxies, that covers 200 $\square ^\circ$ (Mehta et al.  in prep) within the SDSS
Stripe 82 \citep{Annis2014, Jiang2014}.  
Candidates are selected if they show a $g$-band flux excess, that we assume is a result of a strong emission line.  As part of the spectroscopic followup campaign, we have discovered a highly unusual object, hereafter S82-20.  
This galaxy, at $z = 3.082$, is characterized by a narrow emission line spectrum (FWHM $\approx$ 700km s$^{-1}$ ), showing a strong \lya emission, and additional lines from the \civ $\lambda\lambda$ 1548, 1550 \AA\ doublet, \heii $\lambda$ 1640 \AA, \ciii] $\lambda\lambda$ 1907, 1909\AA\ doublet, and, most remarkably, the \ovi $\lambda\lambda$ 1032, 1038 \AA\ doublet.  The observed line ratios cannot distinguish between AGN and SF dominated excitation,
although photoionization by stars is highly unlikely given the presence of the \ovi\ lines that need $\approx$10 Rydbergs to excite.  
Additionally, the galaxy is not detected at long wavelengths ($\lambda >2\mu$m rest-frame), arguing against the obscured Type~II QSO origin.

This paper is organized as follows.  
The data and data reduction, the measurements, and the analysis of the photometric and spectroscopic data are presented in Section \ref{sec:data}.  
In Section \ref{sec:models}, we present the calculation of the emission line ratios from photoionization and shocks models.  
Section~\ref{sec:results} presents the results which are discussed in Section \ref{sec:discussion}.  
The conclusions of this work are presented in Section \ref{sec:conclusion}.
Thorough out the paper, we assume the Lambda cold dark matter ($\Lambda$CDM) cosmology with the following parameters:
$H_0=$70 km s$^{-1}$ Mpc$^{-1}$; $\Omega_m$=0.3; $\Omega_{\Lambda}$=0.7.

\section{Observations and data analysis}\label{sec:data}
In this section, we describe the data available on S82-20.  Section~\ref{sec:spectro} presents the spectroscopic data and the data reduction.  Section ~\ref{sec:measurement} presents the measurements of the emission line fluxes, equivalent width, and velocity width.  Section~\ref{sec:imag} describes available archival imaging in the optical through far-IR as well as the measurements of S82-20 total fluxes (or upper limit) at various wavelengths.
 
\subsection{Spectroscopic Observations}
\label{sec:spectro}
S82-20 [RA=23:48:11.86, DEC=-00:11:47.85 (357.04941596, -0.19661512)] was observed in three occasions.  A first confirmation spectrum in 2013 with the 200-inch Hale Telescope, a second higher resolution spectrum with the 11~m Southern African Telescope \citep[SALT, ][]{SALT} in 2018, and finally a full spectrum extending the rest-frame coverage to the \ciii] lines, using the Large Binocular Telescope (MODS/LBT), in 2020.  The following sections describe the data and data reduction.

\subsubsection{Palomar observations}
A first optical spectrum was acquired with the Double-Beam Spectrograph \citep[DBSP,][]{dbsp} mounted on the Hale 5m telescope at Palomar
Observatory in October 2013.  We used the D55 dichroic to split the light at $\sim 5500$\AA\ into a blue and a red channel.  The red spectrum, acquired with the
600/10000 lines mm$^{-1}$ grating, covers the wavelength range between 6000\AA\ and 8500\AA.  The blue spectrum was acquired with the
600/4000 lines mm$^{-1}$ grating, and covers the wavelength range between 3500\AA\ and 5500\AA.  We used a 1\farcs5 wide slit, resulting in a spectral resolution of 8\AA\ (i.e., $\sim$ 500 km s$^{-1}$) at 5000\AA.  We obtained two exposures of 1,800s each, with seeing full width half maximum (FWHM) of $\sim$ 1\farcs0.  

We performed the basic steps of the data reduction (bias subtraction, flat-field normalization, wavelength calibration, and spectral extraction)
using standard IRAF packages \citep{Tody1986}.  The observing night was clear, but not photometric.  Therefore, we removed the wavelength-dependent instrumental 
response using the sensitivity curve derived from the spectrum of the standard star BD+17\_4708 observed during a photometric night in the same observing run.  
The absolute flux calibration was then performed normalizing the red and blue spectra to the total SDSS $r$-- and $g$--band magnitudes, respectively.  

\begin{figure*}
	\includegraphics[width=0.95\textwidth]{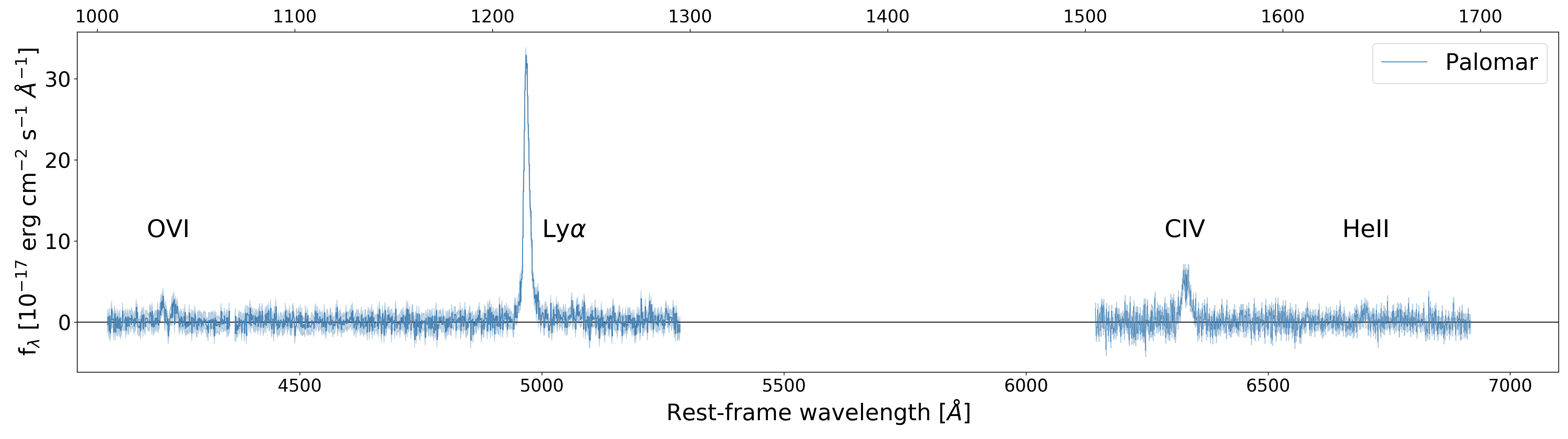}
	\caption{Spectrum of S82-20 obtained with the Hale 5m telescope.  In addition to \lya, the spectrum shows a clear detection of the \civ\ and the \ovi\ doublets.
	}
	\label{fig:palomar}
\end{figure*}

The final extracted Hale spectrum is shown in Figure~\ref{fig:palomar}.

\subsubsection{SALT observations}

\begin{figure*}
	\includegraphics[width=0.95\textwidth]{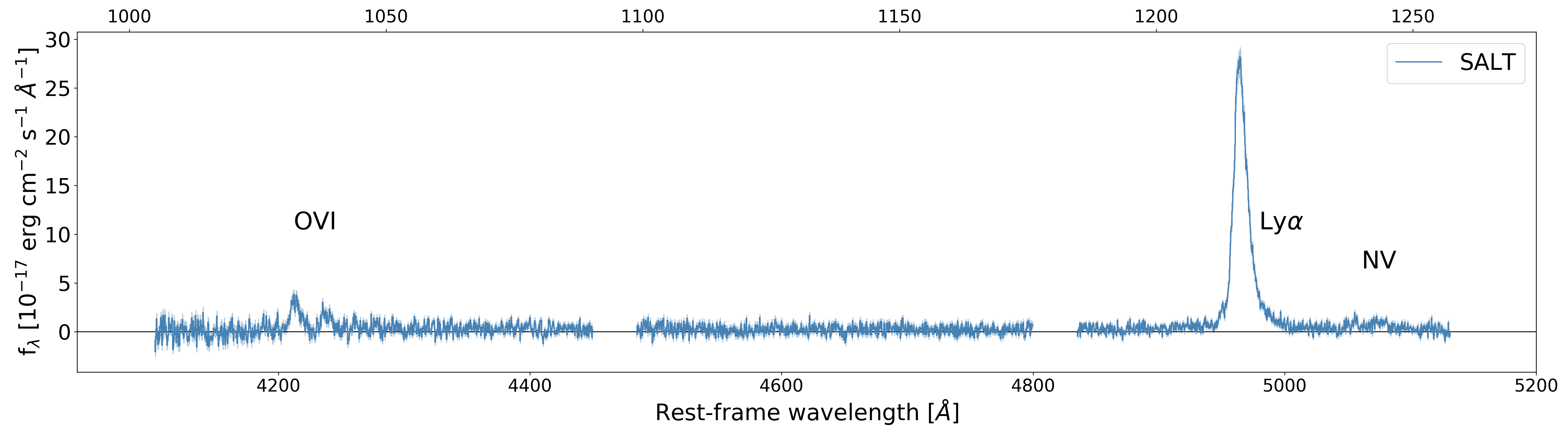}
	\caption{Spectrum of S82-20 obtained with SALT telescope.  Note that the different setup provides higher resolution but a shorter wavelength range than in Figure~\ref{fig:palomar}
	}
	\label{fig:SALT}
\end{figure*}

S82-20 was observed in service mode using the Robert Stobie Spectrograph (RSS, \citet{RSS}) at the SALT telescope.  Observations were acquired between June and September, 2018, over 6 nights.  Data from each night included a 1064 second exposure in long slit mode, with a slit width of 1$\farcs$5.  
The slits were oriented to cover both the target object S82-20, as well as an alignment star, which was later used to flux calibrate the spectrum.
The detector of the RSS spectrograph \citep{RSS} consists of a mosaic of three CCDs, with pixel size of 15 $\mu$m, corresponding to a spatial resolution of 0\farcs26 per pixel using the 2$\times$2 binning.  The seeing FWHM was $\sim$ 1\farcs5 during the observations.  We used the RSS PG2300 grating with a tilt angle 32.375 degrees which provides the wavelength coverage from 4100\AA\ to 5140\AA, with a spectral resolution of 2500.

The basic data reduction steps of overscan, bias and gain correction, crosstalk, mosaicking of the CCDs, and merging of the amplifiers were performed using the SALT primary data reduction pipeline  \citep{pysalt}{\footnote{http://pysalt.salt.ac.za/}}.  Further analysis, including flat-fielding, illumination, wavelength calibration, background sky subtraction, removal of cosmic rays, atmospheric extinction correction, and flux calibration, was performed using the IRAF package.  
The background was subtracted by fitting a second-order polynomial to each column along the cross-dispersion direction.  
We removed the cosmic ray hits using the L.A.Cosmic{\footnote{http://www.astro.yale.edu/dokkum/lacosmic/}} spectroscopic version on IRAF.
After correcting each exposure for atmospheric extinction, we combined them into a single image to increase the signal to noise ratio.  
\\

We extract S82-20 and the reference star separately, using apertures 3\farcs0 and 6\farcs3 wide, respectively.  
To perform the flux calibration of the SALT spectrum we used the reference star observed in the same slit.  This star was observed spectroscopically as part of the SDSS survey in February 2008, and the spectrum is available from Data Release~14.  We derived the wavelength dependent instrument sensitivity function by comparing the RRS and SDSS spectra of the reference star.  We checked that the reference star is not variable using data from Pan-STARRS \citep{PS1}.

The final extracted Hale spectrum is shown in Figure~\ref{fig:SALT}.

\subsubsection{LBT observation}

We observed S82-20 with the Multi-Object Double Spectrographs \citep[MODS,][]{MODS} mounted on the Large Binocular Telescope (LBT) on October 12th, 2020 for a total observing time of 1 hour.  The one-hour integration time was split into three exposures of 1200 seconds each.  
The data were taken using MODS1 and MODS2 in long-slit dual beam G400L/G670L grating mode, using a 1$\farcs$0 wide slit.  The resulting spectrum covers the wavelength range from 3200\AA\ to 10,000\AA, with a resolving power of R=2000.  The observing conditions were clear and the seeing FWHM was $\sim$ 0\farcs9 during the observations.  

We performed the primary detector calibration including bias, flat field and fixing of the bad columns using the public python program modsCCDRed \citep{modsCCDRed}.  The following steps (wavelength calibration, sky subtraction, spectral extraction, and flux calibration) were performed with the modsIDL pipeline \citep{modsIDL}.  The spectrum was extracted using a 2\farcs5 aperture.  To derive the wavelength dependent instrument response we used the spectrum of the standard star BD+28 4211 observed during the same night.  The absolute flux calibration was then performed normalizing the red and blue spectra to the total SDSS S82-20 $r$-- and $g$--band magnitudes, respectively.  We combined the MODS1 and MODS2 spectra via a weighted sum, with the weights equal to the variance of the continuum around the \lya emission for the blue channel and around \civ\ emission for the red channel.  

\begin{figure*}
	\includegraphics[width=1\textwidth]{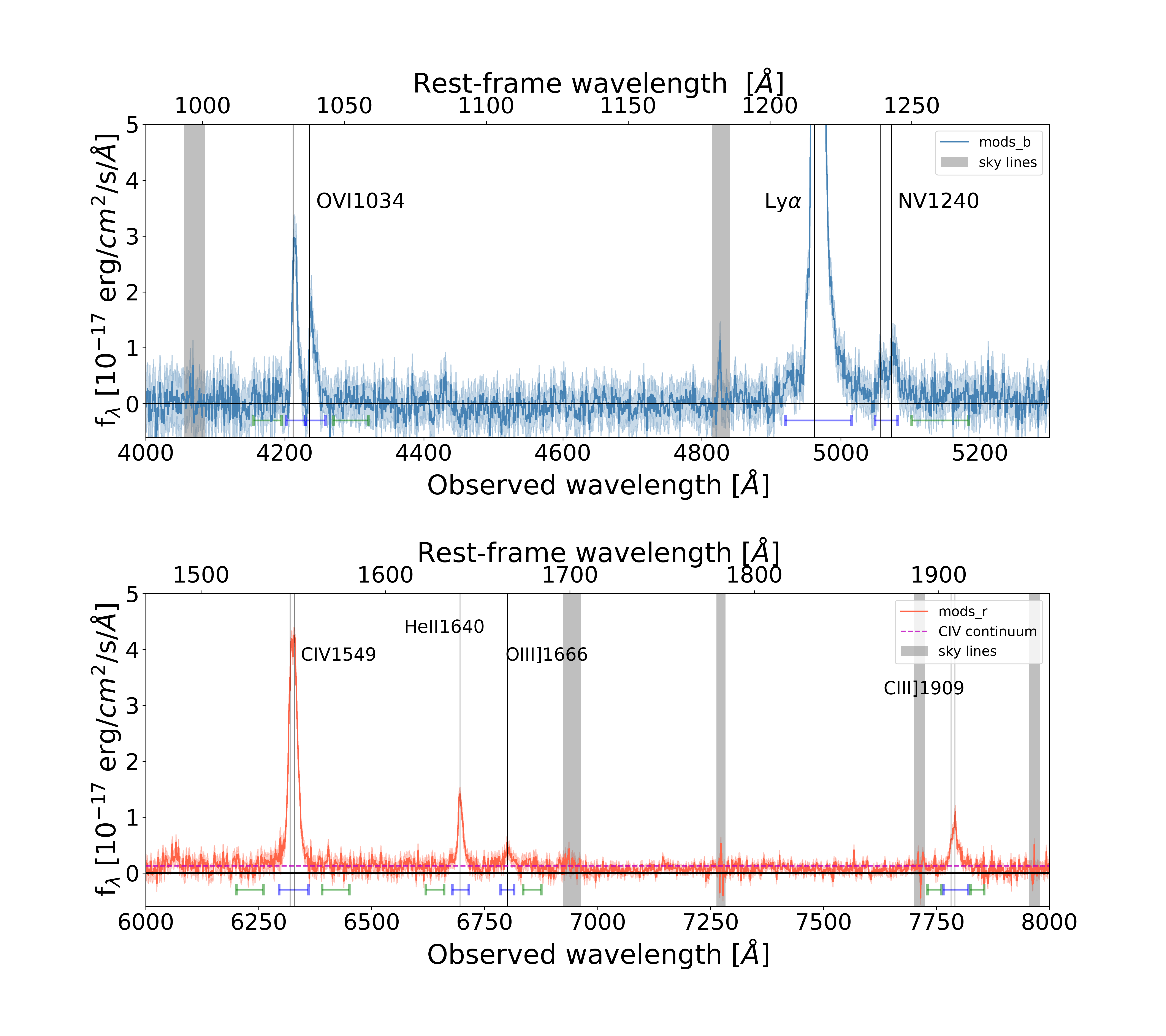}
	\caption{Spectrum from MODS/LBT.  The main emission lines identified are marked.  The blue bars are the integration range for emission line flux, and the green bars are the wavelength range for emission line continuum level.  }
	\label{figmods}
\end{figure*}

The final extracted MODS/LBT spectrum is shown in Figure~\ref{figmods}.

\begin{table}
\caption{S82-20 emission line properties }
 \begin{threeparttable}
	\begin{tabular}{cccc}\hline
		& SALT & Palomar & LBT \\
		\hline
		Line			&	Luminosity & Luminosity & Luminosity \\ 
		& 	10$^{43}$ erg/s		& 10$^{43}$ erg/s 	& 10$^{43}$ erg/s \\ 
		\hline
		\lya 			&	35.45 $\pm$ 0.19 & 	39.45 $\pm$ 1.12	& 36.10 $\pm$ 0.21	\\
		\ovi\ &	3.95 $\pm$ 0.28  &	3.7 $\pm$ 0.1		& 3.70 $\pm$ 0.17	\\
		\nv\ 			&	1.23$\pm$ 0.13   &	$<$1 			& 1.34 $\pm$ 0.12	\\
		\civ\	&	-		   &	10.0$\pm$1.0		& 8.67$\pm$0.05	\\
		\heii\ 			&	-		   &	1.6$\pm$0.2		& 1.52 $\pm$ 0.04	\\
		\oiii]\ 			&	-		   &		-		& 0.38 $\pm$ 0.04	\\
		\ciii]\ 			&	-		   &	 	- 		& 2.04 $\pm$ 0.07	\\
		\hline
		Line & FWHM & FWHM & FWHM\\
		& km s $^{-1}$ & km s$^{-1}$ &	km s$^{-1}$ \\
		\hline
		\lya 			&	734.8$^{+8.5}_{-12.8}$		& 741$^{+21}_{-10}$	& 	774.2$^{+4.1}_{-8.5}$\\
		\ovi\			&	662.0$\pm$42.8		& 726$\pm$88		& 	635.8$\pm$17.9\\
		\civ\			& 	- 				& 819$\pm$85		& 	838.0$\pm$10.7\\
		\heii\			& 	-				&	-		&	575.9$\pm$14.6\\
		\ciii]\			& 	-				&	-		&	775.9$^{+174.7}_{-122.3}$.\\
		\hline
		Line & EW$^a$ & EW$^a$ & EW$^a$ \\ 
		& \AA\	& \AA\ & \AA\ \\
		\hline
		Ly$\alpha$ &	 	-	&	- & 754.3$\pm$165.2 \\
		Ly$\alpha$$^b$ &	 	275.15 $\pm$ 21.8 	&	306.2 $\pm$ 27.5 & 280.3 $\pm$ 22.2 \\
		\ovi$^b$ &		30.63 $\pm$ 3.2 	&	28.7 $\pm$ 2.2 & 28.7 $\pm$ 2.6 \\
		\nv$^b$  & 			9.5 $\pm$ 1.3	& 	- 		& 	10.4 $\pm$ 1.2	\\
		\civ\	&	-	&	229.6 $\pm$ 44.8$^c$ &	192.4 $\pm$ 16.7	\\
		\heii\	&	-	& 	36.7 $\pm$ 7.4$^c$ 	& 	29.7 $\pm$ 2.5		\\
		\ciii]\	&	-	&	-	&  38.1 $\pm$ 4.2	\\
		\hline
	\end{tabular}
	\label{tab:lines}
  \begin{tablenotes}
		\item{$^{\rm a}$Rest-frame equivalent width.}
		\item{$^{\rm b}$Continuum derived from $g$ band imaging.}
		\item{$^{\rm c}$Continuum derived from $r$ band imaging.}
  \end{tablenotes}

 \end{threeparttable}
\end{table}

\subsection{Emission line measurements}\label{sec:measurement}
The first spectrum obtained at the Hale telescope confirms that the color excess in the $g-$band is due to a strong \lya emission line, and that the object is at $z=3.082$.  In addition to the detection of strong \lya, the Hale spectrum reveals the presence of high ionization lines, including the \civ$\lambda\lambda$1549, 1551\AA\ (hereafter \civ\ doublet) and \ovi$\lambda\lambda$1032, 1038\AA\ (hereafter \ovi\ doublet).  The spectrum additionally indicates the presence of \heii\ and \nv\ emission lines, although the signal-to-noise ratio in these faint lines is low.
The SALT spectrum confirms the detection of the \ovi\ doublet, but does not extend red enough to cover the \civ\ doublet and \heii\ lines.  

In addition to the bright \ovi, \lya, and \civ\ lines, the LBT spectrum, shown in Figure~\ref{figmods}, clearly reveals the presence of \heii$\lambda$1640 (hereafter \heii), \oiii]$\lambda$1666 (hereafter \oiii]), and the \ciii]$\lambda\lambda$1907,~1909 doublet (here after \ciii]).  
We measured the flux of the emission lines by directly integrating the spectral flux density over the wavelength range of the lines shown in Figure~\ref{figmods}, after subtracting the continuum.  We derived the continuum for each line by averaging the flux density over the spectral regions indicated in Figure~\ref{figmods}.  For most lines we use two 25\AA-wide regions, one on each side of the emission line.  For \lya, in order to minimize biases introduced by the \lya forest, we only measure the continuum on the red side of the line.  Even though we find that the continuum is not detected at more than 3$\sigma$, we properly account for its contribution in the line flux error calculation.  

For the emission lines in which the continuum was not detected in the spectrum (see Table~\ref{tab:lines}), we compute the observed equivalent width (EW) using the continuum derived from the emission-line-corrected broad band fluxes as follows.  Assuming that the $g$ and $r$ bands are dominated by the \lya $+$\ovi\ and \civ\ emission lines, respectively, we can write the continuum flux density ($f^{cont}_{\lambda}$) as:

\begin{equation}
f^{cont}_{\lambda} =f_{g,r}- \frac{F_{\rm line}}{W_{g,r}}, \\
\end{equation} 

\noindent 
where $f_{g, r}$ are the flux density computed from the $g$ and $r$ band magnitudes, $W_{g,r}$ are the FWHM of the $g$ and $r$ band filters, and $F_{\rm line}$ is the total flux of the emission line.  
We measure continuum levels of $f^{cont}(g)=3.6\pm 0.3 \times 10^{-18}$erg cm$^{-2}$ s$^{-1}$ \AA$^{-1}$ and $f^{cont}(r)= 1.2 \pm 0.2\times 10^{-18}$erg cm$^{-2}$ s$^{-1}$ \AA$^{-1}$.  These values are consistent with the upper limits derived from the spectra.

Finally, to compute the FWHM of the emission lines we proceed in two different ways for \lya and the remaining, weaker lines .  For \lya we compute the FWHM numerically, by finding the width of the profile where the value is half of the maximum value.  The error on the FWHM is then computed via a Monte Carlo simulation.  We generate 1000 realizations of the \lya profile by randomizing the spectrum within $\pm \sigma_{cont}$ at each wavelength.  For each realization, we compute the line FWHM, and then estimate the error as the standard deviation of the distribution of the 1000 realizations.  
For the lower S/N emission lines, we measure the FWHM by fitting a Gaussian profile to the observed spectra.  
The \civ\ and \ovi\ doublets have the intrinsic 2:1 line ratio in the low electron density and optical thin environment \citep[$n_e\tau_0 < 10^{16} $cm$^{-3}$,][]{Hamann1995, NIST_ASD}.
Since we measure the \ovi\ doublet line ratio $\simeq$ 2:1, we fix the \civ\ doublet line ratio as 2:1 and fit with a double Gaussian profile.  
The flux ratio of \ciii]1907/\ciii]1909 varies from 1.53 to 0 with electron density \citep{Keenan1992}.  
While the line profile in our spectrum is not clear to distinct the doublet.  
We generate 10,000 realizations of the \ciii] profile by randomizing the spectrum within $\pm \sigma_{cont}$ at each wavelength.
The FWHM of \ciii] by fitting a single Gaussian profile is 775.9$^{+174.7}_{-122.3}$ km s$^{-1}$.

Table~\ref{tab:lines} presents the measurements performed on the available spectra.  For the doublets (\ovi\, \civ\ and \ciii]) we report the sum of the two lines in the doublet.  In the following analysis, we use the emission line fluxes measured on the LBT spectra, unless we state otherwise.  


\subsection{Photometric measurements}
\label{sec:imag}

\begin{figure*}
	\includegraphics[width=1\textwidth]{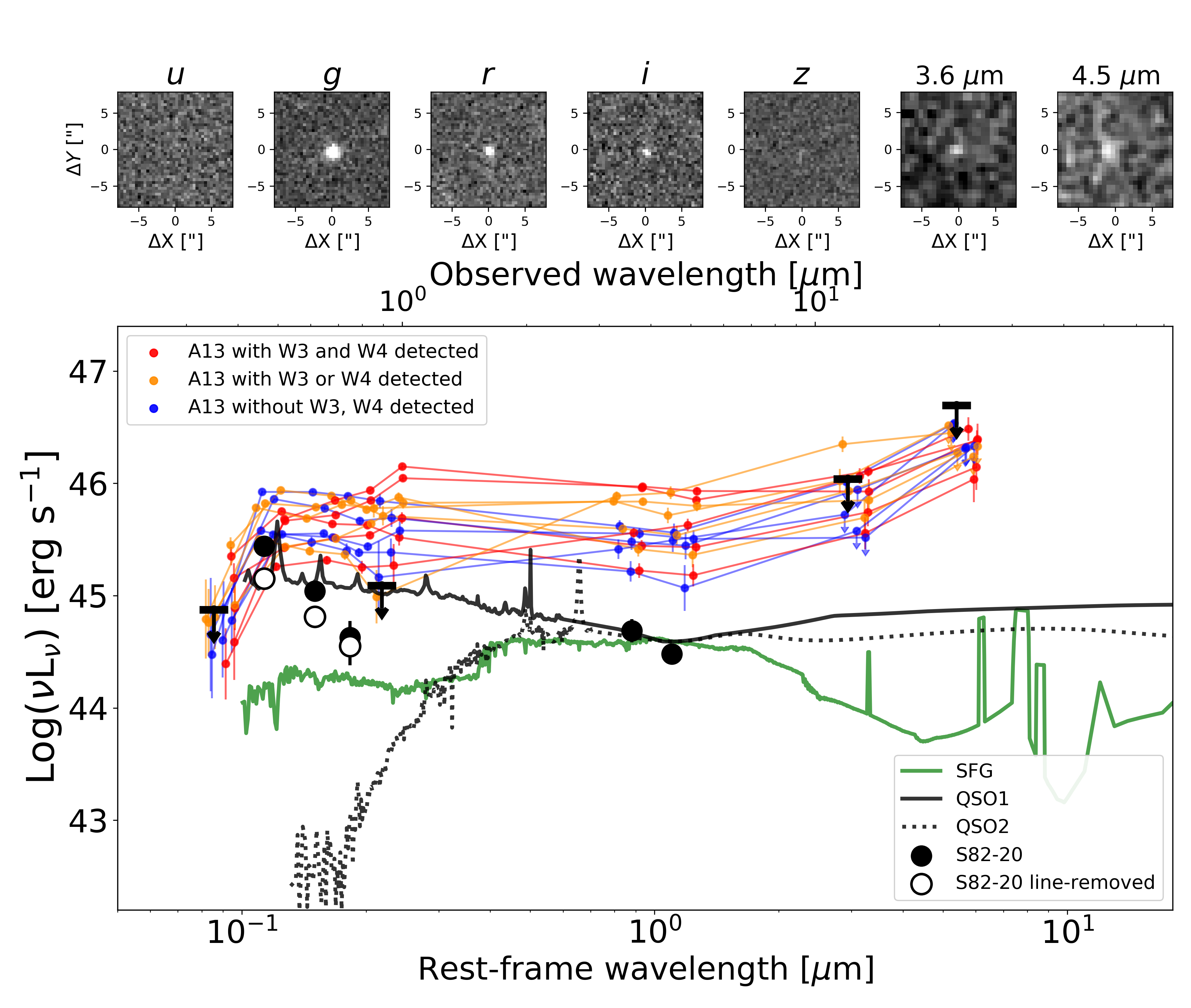}
	\caption{The top panel shows the images of S82-20 in SDSS u, g, r, i, z band and IRAC 3.6 $\mu$m, 4.5$\mu$m.  The bottom panel shows the spectral energy distribution (SED) of S82-20 (large points, and upper limits) compared with spectral templates of QSOs (both Type I and Type~II) and a star-forming galaxy by \citet{Polletta2007}.  We also show for comparison the observed SEDs of the A13 Type~II QSO candidates, distinguishing between those with both (red points), either one (orange points), and without (blue points) detection at wavelength longer than 6$\mu$m.  }
	\label{figsed}
\end{figure*}

The field of S82-20 was imaged at optical to mid-infrared wavelengths with a number of instruments.  
The measurements of the total flux in the different bands, described below, are presented in Table~\ref{tab:photometry}.  We show the resulting spectral energy distribution in Figure~\ref{figsed}.  

At the optical wavelengths, the field was observed as part of the SDSS Stripe~82 campaign \citep{Annis2014}.  The fluxes in the $u,\,g,\,r,\,i$ and $z-$ bands were measured using circular apertures of 3\farcs56 radius.  In Figure~\ref{figsed} we show both the direct measurements (filled circles), as well as the values corrected for the contribution of the emission lines (open circles).  The $g-$ and $r-$band are clearly affected by the presence of \lya and \civ, respectively.
At longer wavelengths, the field of S82-20 is covered by the Spitzer-IRAC Equatorial Survey (SpIES) program $3.6 \mu$m and $4.5 \mu$m \citep{Timlin2016}, by the Wide-field Infrared Survey Explorer at 12, and 22$\mu$m \citep{Wright2010}, and by the Herschel telescope at 250, 350, and 500$\mu$m as part of the HerMES Large Mode Survey (HeLMS) Survey \cite[][]{Asboth2016}.

The co-added Spitzer images were downloaded from the IRSA archive and the fluxes were calculated within circular apertures.  The object is clearly detected in both Spitzer bands as shown in the top panels of Figure~\ref{figsed}.  To compute the total flux in the IRAC bands we used circular apertures with a $2\farcs4$ radius.  The background was evaluated within a circular annulus with inner radius of 7\farcs2 and outer radius of 12$''$ and subtracted from the total flux measured within the aperture.  

We used the WISE/NEOWISE Coadder\footnote{https://irsa.ipac.caltech.edu/applications/ICORE/} tool to create a deeper co-add from all 15 available single-exposure images covering S82-20 \citep{Masci2009}.  S82-20 is not detected in the WISE images and in Figure~\ref{figsed} we show the 3$\sigma$ upper limits.  The upper limits were computed within the circular aperture of 8\farcs25 and 16\farcs5 radius, at 12 and 22$\mu$m, respectively, to account for the wavelength dependent instrumental PSF.

We inspected the HeLMS data at the position of S82-20 and find that the object is not detected in any of the maps, implying 3$\sigma$ upper limits to the flux of 15.6, 12.9, and 10.5 mJy at 250, 350, and 500 $\mu$m, respectively.  Finally, the object is not detected in the Faint Images of the Radio Sky (FIRST) project that observed the area at a frequency of 1.4~GHz using the NRAO Very Large Array (VLA).
The non detection corresponds to an upper limit on the luminosity of 2.8$\times 10^{29}$W/Hz.

\begin{table}
\centering
  \caption{Photometric measurements}
 \begin{threeparttable}
		\begin{tabular}{lc}
			Filter  & $m_{AB}$    \\
			\hline
			SDSS u  & $>$23.89   \\
			SDSS g  & 22.17$\pm$0.05 (22.9$\pm$0.1)$^a$\\
			SDSS r  & 22.87$\pm$0.10 (23.4$\pm$0.2)$^a$ \\
			SDSS i  & 23.68$\pm$0.37 (23.9$\pm$0.4)$^a$ \\
			SDSS z  & $>$22.33   \\
			IRAC 3.6 & 21.83 $\pm$ 0.26 \\
			IRAC 4.5 & 22.10 $\pm$ 0.18 \\
			WISE W3 & $>$17.1   \\
			WISE W4 & $>$14.8 \\
			\hline  
		\end{tabular}
		\label{tab:photometry}
  \begin{tablenotes}
	\item{$^{\rm a}$The values in parenthesis are corrected for the contribution of emission lines.}
	\end{tablenotes}
 \end{threeparttable}
\end{table}

\section{Modeling of line ratios}\label{sec:models}
The LBT spectrum of S82-20 allows us to compute a number of line ratios using rest-frame UV emission lines.  These line ratios can be compared with photoionization and shock models to identify the main excitation source responsible for the large \lya luminosity.  In this section we briefly describe the details of the models used to interpret the line ratios in Section~\ref{sec:discussion}.

\begin{table*}
	\centering
	\footnotesize
	\caption{ Adjustable parameters of the photoionization models.  }
	\begin{tabular}{ccc}
		\hline
		\hline
		Parameter & AGN  & SFG  \\
		\hline
		Ionizing spectrum  & $\alpha=-1.7$  & m$_{up}$=100 M$_\odot$  \\
		log($U_s$) &  -1.0, -2.0, -3.0, -4.0, -5.0  &  -1.0, -2.0, -3.0, -4.0  \\
		log(n$_H$/cm$^{-3}$) &  2, 3, 4 & 2, 3  \\
		Z & 0.0001, 0.001, 0.008, 0.017, 0.03 & 0.0001, 0.001, 0.008, 0.017, 0.03 \\
		$\xi_d$ & 0.3 & 0.3 \\
		\hline
	\end{tabular}
	\label{tab:model}
\end{table*}

\subsection{AGN NLR model}
We use the most up-to-date grid of the AGN NLR photoionization models of \citet{Feltre2016}, presented in \citet{Mignoli2019} .  For the star-forming galaxies, we used the models of \citet{Gutkin2016}.
Both models use the CLOUDY photoionization code \citet[][]{Ferland2013}.  
\\

The AGN models assume that the gas has an open geometry, appropriate when the covering factor of the gas is small.  The AGN ionizing spectrum is described with a series of broken power laws: 

\begin{equation}
S_\nu=
\begin{cases}
\nu^\alpha & 0.001\le \lambda/\mu m\le 0.25, \\
\nu^{-0.5} & 0.25\le \lambda/\mu m\le 10.0, \\
\nu^2 & \lambda/\mu m > 10.0,
\end{cases}
\label{eq:2}
\end{equation}

\noindent
where $S_\nu$ is the luminosity per unit frequency of the accretion disc.  We assume $\alpha=-1.7$ consistent with the value derived by \citet{Lusso2015} in a stacked UV spectrum of 53 luminous quasars at z$\sim$ 2.4.  We keep the AGN accretion-disk bolometric luminosity fixed at $L_{AGN}=10^{45}$ erg s$^{-1}$.  The model assumes that the NLR has an inner radius of $r_{in}$=90 pc, and the internal microturbulence has a characteristic velocity $\nu_{micr}=$100 km s$^{-1}$.  We assume a dust-to-heavy element mass ratio, $\xi_d$ of 0.3.

\subsection{SFG model}
For the star-forming galaxies, we used the models of \citet{Gutkin2016}, computed following the approach outlined by \citet{Charlot2001} and using an updated version of the \citet{Bruzual2003} stellar population synthesis model (described in section 2.1 of \citet{Gutkin2016}).  A closed spherical geometry is used, which is more appropriate to describe the physical conditions of ionization bounded \hii\ regions.  
We consider the models computed for a standard \citet{Chabrier2003} initial mass function truncated at 0.1 and 100~M$_\odot$, a C/O abundance ratio =1(C/O)$_\odot$ ($\approx$ 0.44), constant star formation rate and age of 100~Myr.  We assume the same dust-to-heavy element mass ratio as in the AGN model, $\xi_d$=0.3.  

In Figure~\ref{figlineratio} we show different tracks for both AGN and SF-galaxies, corresponding to different values of the ionization parameter at the edge of the Str{\"o}mgren sphere ($U_s$), the hydrogen gas density, and the interstellar (gas + dust) metallicity ($Z$).  The parameters varied in the tracks are summarized in table~\ref{tab:model}.

\subsection{Shock model}
We also consider the nebular emission produced from radiative shocks computed by \citet{Alarie2019} using the latest shock and photoionization code MAPPING~V \citep{Sutherland2017}.  The main adjustable parameters in the shock models are the shock velocity (that ranges from 100 to 1000 km s$^{-1}$), the metallicity of the gas with the same sets of element abundances as adopted in the SFG and AGN models \citep{Gutkin2016} described above, the pre-shock gas density (from 1 to $10^4$ cm$^{-3}$), and the transverse magnetic field (from 10$^{-4}$ to 10 $\mu$G).  
We chose the same C/O ratio adopted for the AGN and star-forming galaxies photoionization models (C/O)$_\odot$=0.44).

The pre-shock density and the transverse magnetic field have less impact on the UV line ratios we study here, compared to the shock velocity and metallicity.  Therefore, for simplicity, we fix the pre-shock density to 1~cm$^{-3}$, and the transverse magnetic field to 0.5 $\mu G$.  In Figure~\ref{figlineratio} we show the line ratios from both the precursor and shocked gas.

\section{Results}\label{sec:results}

The optical spectra presented in Figure~\ref{fig:palomar} to~\ref{figmods} confirm that S82-20 is a high-redshift object ($z\sim 3.1$), identified in our survey because of the excess flux in the $g-$band caused by a strong \lya emission line.  In what follows, we present the measurements performed on the LBT/MODS spectrum (shown in Figure~\ref{figmods}), as it is the deepest of the three and covers the longest wavelength range (rest-frame 1000--1950\AA), unless we say otherwise.

The integrated flux of the \lya emission line corresponds to a luminosity of $(3.61\pm0.02) \times 10^{44}$~erg~s$^{-1}$, placing S82-20 among the most luminous \lya emitters known \citep{Marques-Chaves2020LAE}.  These luminosities are commonly associated with AGN activity \cite[see, e.g.,][]{Matthee2017, Sobral2018, Calhau2020} and/or powerful \lya nebulae \citep[e.g.,][]{Steidel2000,Scarlata2009,Herenz2020}, although exceptionally \lya--bright star-forming galaxies have recently been found \citep[see, e.g.,][]{Marques-Chaves2020}.  

It is unlikely that the large \lya luminosity is produced in an extended \lya-nebula.  The broad $g-$band image containing the \lya line reveals that the morphology of S82-20 is compact (see Figure~\ref{figsed}) and spatially unresolved in the seeing limited images.  This morphology rules out the \lya halo interpretation, as these objects have \lya sizes of $\approx$ 100~kpc, and are clearly resolved in ground based imaging \citep[e.g.,][]{Prescott2012}.  

In the rest-frame UV spectrum shown in Figure~\ref{figmods} we identify multiple of high-ionization emission lines.  Redward of \lya, the spectrum shows weak \nv\ and \oiii] lines as well as relatively strong \civ, \heii, and \ciii] lines.  Blueward of \lya, we detect both emission lines of the \ovi\ doublets.  The presence of these highly ionized ions (particularly the detection of the \ovi) suggest the presence of a hard ionizing spectrum, possibly associated with an AGN.  Contribution to the ionization, however, from young stars can not be ruled out, as metal poor dwarf starburst galaxies have been known to show high ionization emission lines of \civ\ and \heii\ \citep[e.g.,]{Stark2015, Mainali2017, Berg2019, Senchyna2019}.  
 
All lines with sufficient signal-to-noise ratio have consistent line widths, with FWHMs of $\approx$750~km~s$^{-1}$.  The FWHM of \civ\ is somewhat larger than that of the other lines, at $838\pm10.1$~km~s$^{-1}$.  The doublet, however, is unresolved at the resolution of the LBT spectrum, so we do not consider this larger FWHM significant.  Corrected for instrumental resolution the measured FWHM of 750~km~s$^{-1}$ corresponds to a velocity dispersion $\sigma = 280$~km~s$^{-1}$.  This value is large, but not impossible in galaxies, particularly at $z>2$ \citep{Lehnert2013, NMF-Schreiber2009}.

The LBT spectrum shows a faint, but detected, continuum redward of the \lya emission line.  The flux density of 1.35$\times 10^{-18}$~erg~s$^{-1}$~cm$^{-2}$~\AA$^{-1}$ corresponds to a continuum luminosity $\log(L_{1300})=41.7\pm0.1$~erg~s$^{-1}$~\AA$^{-1}$, or $M_{UV}=-22.7$, and $10^{11.2}L_{\odot}$.  
The S/N, however, is too low to search for the presence of absorption lines, that could help pin down the origin of the detected continuum.  Combined with the large \lya luminosity, this continuum implies a rest-frame EW of $754.3\pm$ 165.2\AA\ (see Table~\ref{tab:lines}).  This value is within 2.5$\sigma$ from the EW computed using the emission-line-corrected broad band flux (280.3$\pm$22.2\AA).  

\begin{figure}
	\includegraphics[width=0.48\textwidth]{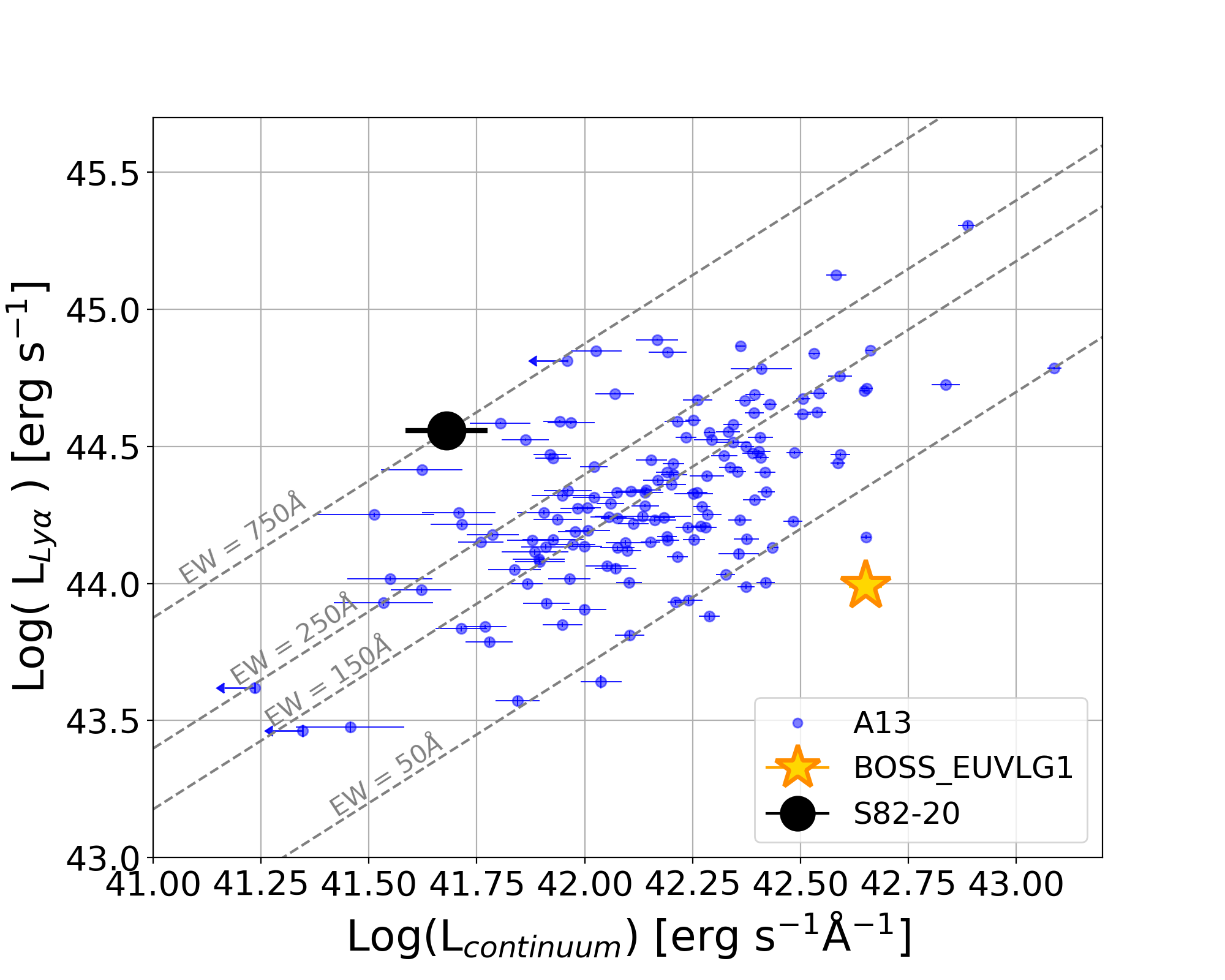}
	\caption{\lya luminosity versus continuum luminosity at 1300\AA, for the A13 sample of Type~II QSOs (blue circles), BOSS-EUVLG1 (yellow star) and S82-20 (black circle).  The position of S82-20 with respect to the A13 sample indicates that it has a larger continuum obscuration compared to the A13 objects.  \label{figLyaCIV}  }
\end{figure}

In Figure~\ref{figLyaCIV} we compare the \lya and continuum luminosity of S82-20 with a sample of luminous \lya sources.  We include both AGN and the only star-forming galaxy known with similar luminosity as our object.  For the AGN, we consider the A13 sample of Type~II QSOs.  This sample was identified by \citet{Alexandroff2013} based on the presence of strong and narrow (with FWHM $<$ 2000 km s$^{-1}$) UV emission lines (their "Class A" sample).  We also show BOSS-EUVLG1, the \lya brightest known star-forming galaxy known \citep{Marques-Chaves2020}.  For consistency, we re-measured the emission line fluxes and continua for the A13 sample, using the original SDSS spectra and the same technique discussed in Section~\ref{sec:data}.  We only consider A13 objects at $z>2.6$, for which the \ovi\ emission line is covered by the SDSS spectra.  For BOSS-EUVLG1 we report the published values, assuming the galaxy has a flat spectrum between 1500 and 1300\AA\ \citep[consistent with the published spectrum, see Figure~1 in][]{Marques-Chaves2020}.  The A13 Type~II QSO candidates show a general positive correlation between the \lya luminosity and the rest-frame UV continuum.  S82-20 is in the upper envelope of this trend, with the UV continuum $\approx$0.7dex fainter than what would be predicted based on its \lya luminosity.  Similarly, BOSS-EUVLG1 is an outlier in this correlation, at a much lower \lya luminosity, given its very bright ($\approx 30 L^*_{UV}$) UV continuum.

\begin{figure}
	\includegraphics[width=0.47\textwidth]{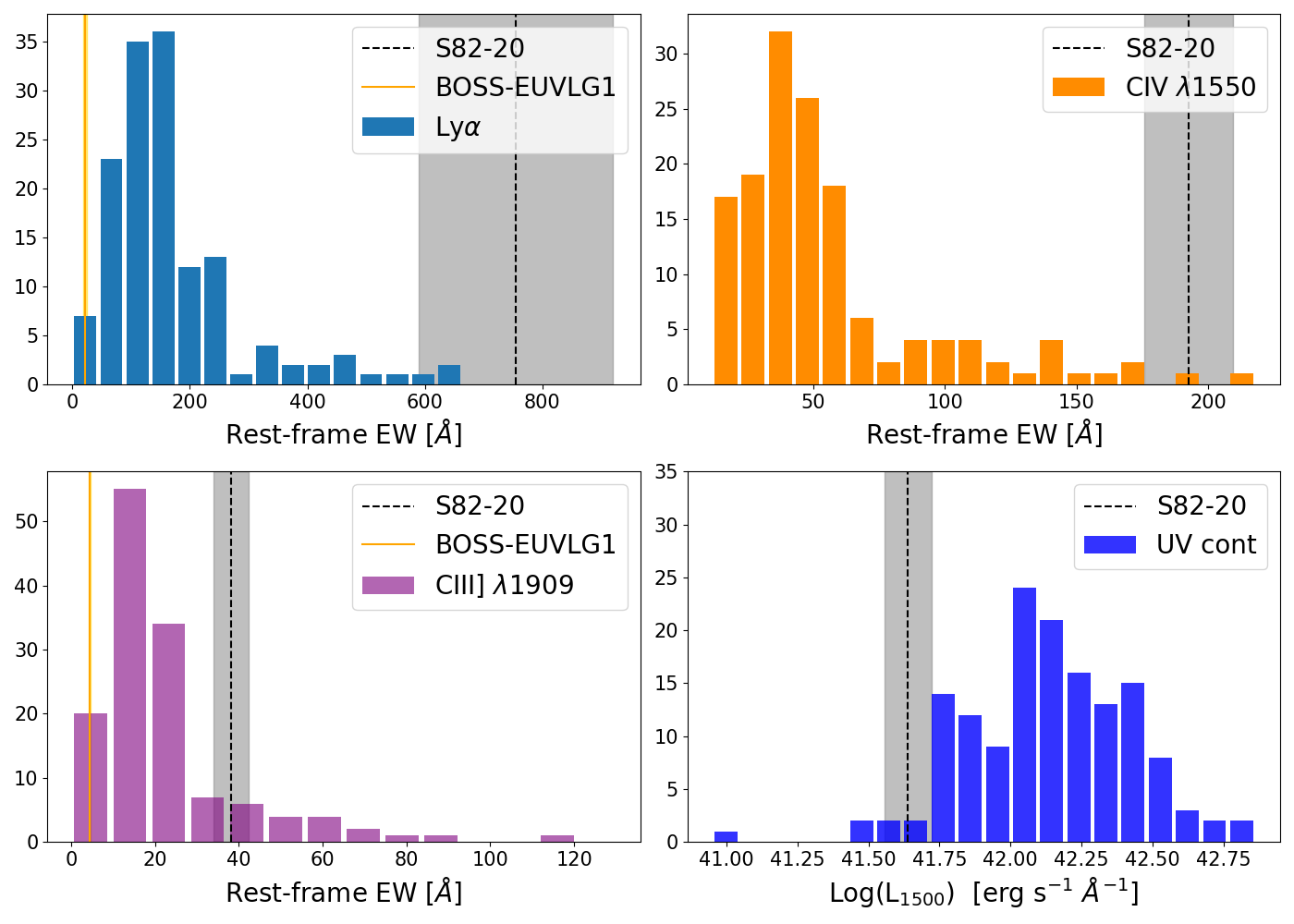}
	\caption{ The rest-frame EW of \lya, \civ, \ciii] and the UV continuum luminosity statistics of S82-20 (dash line and gray area), BOSS-EUVLG1 (yellow line) and Type~2 quasar candidates in \citet{Alexandroff2013}.  }
	\label{fighis}
\end{figure}

\begin{figure*}
	\centering
	\includegraphics[width=1.0\textwidth]{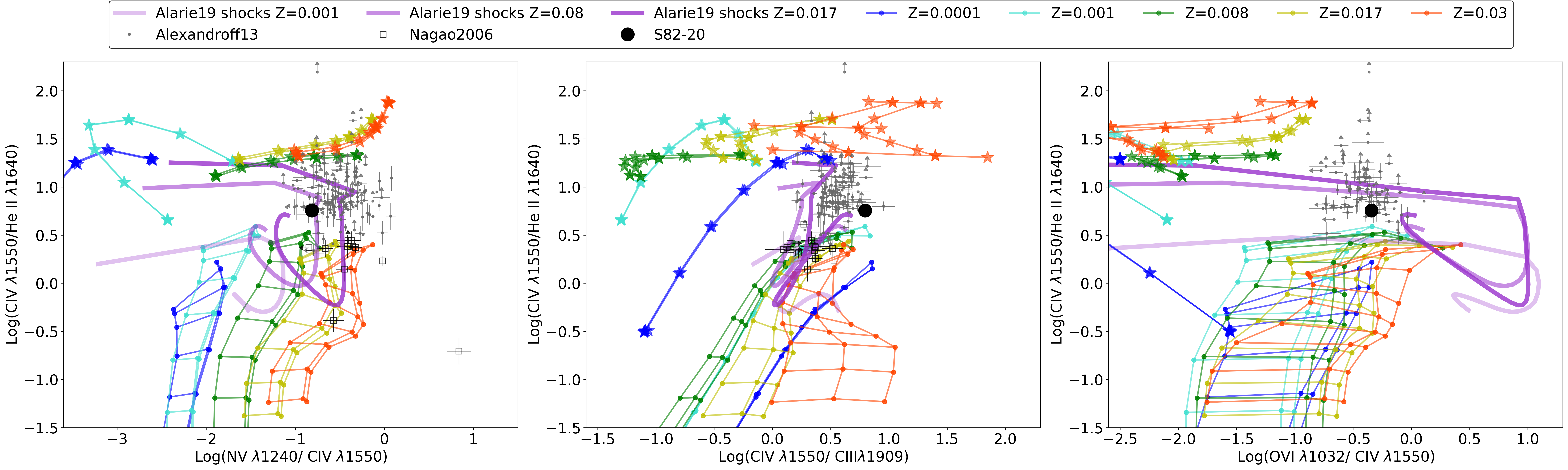}
\caption{Line ratios diagnostic plots for S82-20 and the sample of Alexandroff et al.  (2013).  We show \civ/\heii\ as a function of \nv/\civ\, \civ/\ciii] and \ovi/\civ, in the left, middle and right panel respectively.  We also show line ratios predicted by photoionization models for star-forming galaxies (star markers) from \citet{Gutkin2016} and AGN (dot markers) from \citet{Feltre2016}.  Finally, the purple lines show the line ratios computed with the shocks model from \citet{Alarie2019}.}
	\label{figlineratio}
\end{figure*}

Due to the resonant nature of the transition, \lya can be enhanced/suppressed with respect to the continuum by \emph{ad hoc} geometries of the neutral gas and the presence of even a small amount of dust \citep{Dijkstra2006}.  Therefore, in Figure~\ref{fighis}, we compare the EW of other lines detected in S82-20 with those of the A13 sample and BOSS-EUVLG1.  Clearly, the EW enhancement of S82-20 is present in all UV lines.  The \lya, \civ\ and \ciii\ EWs of S82-20 are larger than 99\%, 98\%, and 85\% of the obscured QSOs, respectively.  The other UV lines identified in BOSS-EUVLG1 behave similarly to the \lya and are in the lowest quartile of the EW distributions.  The bright \lya luminosity and narrow lines would suggest that S82-20 is powered by an obscured QSO, similar to those in the A13 sample.  The continuum, however, is substantially fainter, implying either a higher obscuration or an enhancement of the emission line fluxes compared to the A13 sample.  

In Figure~\ref{figsed} we show the spectral energy distribution of S82-20.  Filled and open circles show the direct measurements and those corrected for the emission line contributions, respectively.  The rest-frame UV is remarkably blue.  The observed emission-line-corrected color of $g-r=-0.5$ (see Table~\ref{tab:photometry}) corresponds to a UV slope $\beta = -3.3^{+0.6}_{-0.9}$, where we followed the convention of parameterizing the continuum as $f_{\lambda}\propto \lambda ^\beta$.\footnote{Written in terms of frequency, the slope corresponds to $\alpha = 2.3^{+1.5}_{-1.4}$ (see Equation~\ref{eq:2}).} This continuum is substantially bluer than the typical Type~I QSOs, as shown in Figure~\ref{figsed} where we compare the SED of S82-20 to a set of templates from the SWIRE library \citep{Polletta2007}.  All templates are normalized to the S82-20's energy at rest-frame 1$\mu$m.  This wavelength was chosen for reference, as it probes a galaxy's stellar mass for objects with either a heavily obscured or low luminosity AGN.  Additionally, the effect of dust is minimal at 1$\mu$m.  

In Figure~\ref{figsed} we also show (as the colored points and connecting lines) the SEDs of the A13 Type~II QSO candidates at redshifts 2.6$<z<$3.2.  The WISE data (covering the observed 3.4 to 22 $\mu$m) for the A13 sample are taken directly from Alexandroff et al.  (2013), while the optical data were retrieved from the SDSS archive.  The SDSS photometry for the A13 objects is contaminated by the presence of emission lines so we compare with the filled circles for S82-20.  At the wavelength of the \lya emission line ($g-$band) S82-20 falls square in the middle of the energy range covered by the A13 sample, suggesting that, if indeed powered by an accretion disk, S82-20 should have similar ionizing power as these Type~II QSOs.  As we move to longer wavelengths, however, the energy of S82-20 drops much faster than in the A13 sample.  At rest-frame $\approx 1.0\mu$m, S82-20 is almost an order of magnitude fainter than the bulk of the continuum energies of the A13 sample.  This trend continues in the near IR, up to rest-frame 1$\mu$m, where again the luminosity of S82-20 is one order of magnitude fainter than the bulk of the A13 objects.  

\begin{figure*}
		\includegraphics[width=0.48\textwidth]{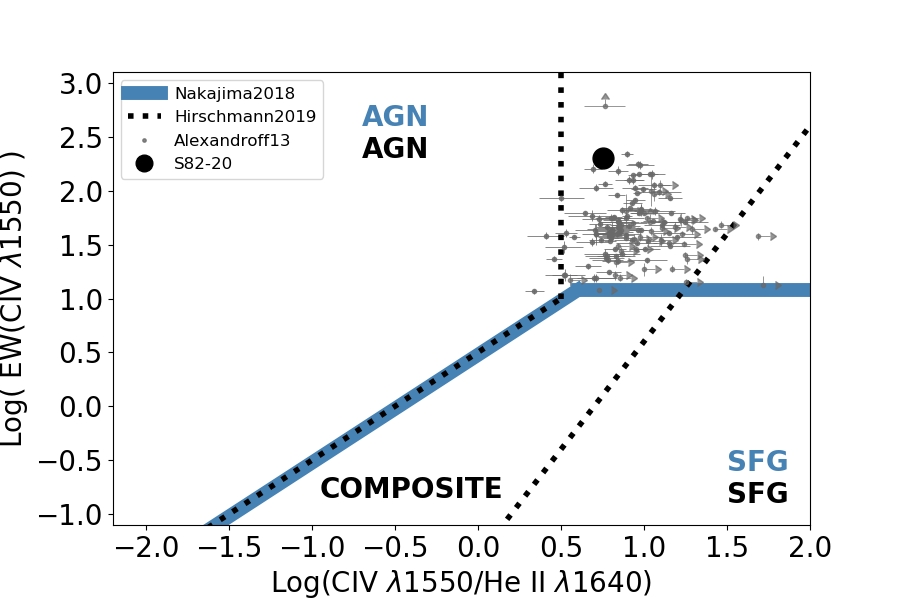}
		\includegraphics[width=0.48\textwidth]{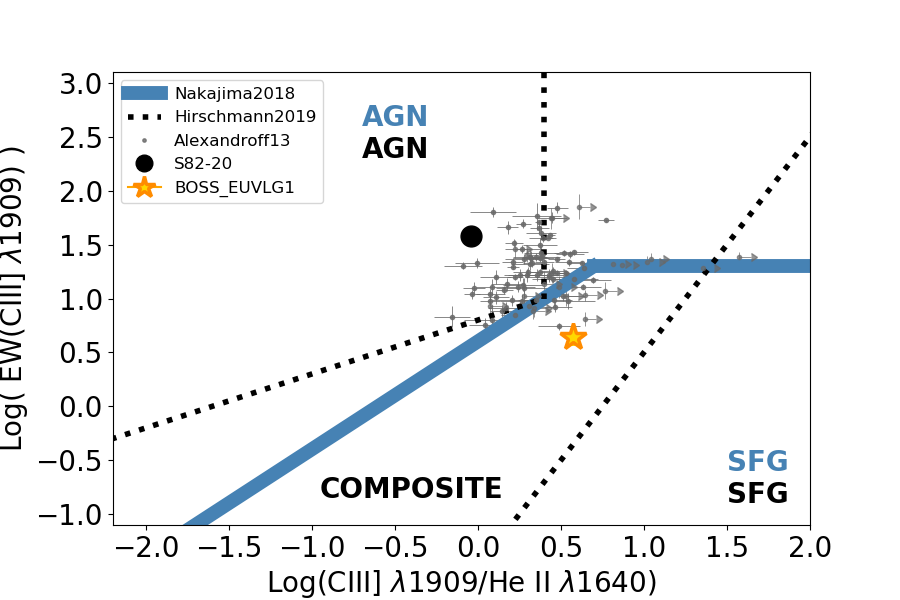}
		\caption{Equivalent width and line ratios diagnostic plots for S82-20 and the sample of Alexandroff et al.  (2013).  We show the EW(\civ) vs \civ/\heii\ and EW(\ciii]) vs \ciii]/\heii\ discriminate lines used in \citet{Nakajima2018} (blue) for star-forming galaxies and AGN, and \citet{Hirschmann2019} (black) for star-forming galaxies, AGN and the composite galaxies.}
	\label{figEWCHe}
\end{figure*}

\section{Discussion }\label{sec:discussion}

The evidence presented in the previous section outline similarities between S82-20 and the population of narrow--line Type~II QSOs.  S82-20 shares many properties with the sample of Alexandroff et al.  (2013), including a similar \lya luminosity, and width of the emission lines.  There are, however, differences that are worth exploring: S82-20 has a substantially \emph{fainter} and \emph{bluer} continuum than the bulk of Type~II QSOs with similar \lya luminosity, and it is not detected at 5$\mu$m rest-frame, where many of the A13 objects are (See Figure~\ref{figsed}).  

The combination of the fainter continuum with similarly bright emission lines suggest that the S82-20 energy source is an intrinsically powerful QSO (at least as powerful as the A13 objects) but more heavily obscured.  In fact, A13 discuss how the obscuration of their objects is not extreme (with $A_V\approx 0.5$), as revealed by the detection of faint broad wings in rest-frame optical permitted emission lines.  
The problem with this interpretation, however, is the non-detection of S82-20 in the WISE bands.  Indeed, the rest-frame 3-6$\mu$m range probed by WISE should be sensitive to emission from hot dust heated by the energy absorbed in the UV.  Energy arguments would suggest that the intrinsic UV luminosity from the accretion disk in the A13 and S82-20 are similar (as they result in the emission of similar \lya luminosities).  Given the inferred larger obscuration, it is then reasonable to expect that S82-20 should present similar re-emitted luminosities as the A13 in the WISE bands, which is not the case.

Another possibility that could explain the large EWs of the emission lines, is that the lines luminosities are enhanced by another mechanism, possibly star-formation or shocked gas.  

\subsection{Type~II Emission line ratios}
In Figure~\ref{figlineratio} we explore the source of ionization of the gas using line ratios computed with the detected rest-frame UV emission lines.  Specifically, we show \civ/\heii\ as a function of \nv/\civ, \civ/\ciii]\ and \ovi/\civ, in the left, middle and right panels, respectively.  S82-20 is marked as a solid black circle, while the A13 Type~II QSOs are shown as gray points.  Figure~\ref{figlineratio} shows that the line ratios for S82-20 are broadly consistent with those of the A13 Type~II QSO sample, although S82-20 tends to have a lower \nv/\civ\ ratio and a higher \civ/\ciii] ratio than the bulk of the A13 emission line selected Type~II QSOs.  In the left and middle panels, we also show the position of a sample of narrow-line X-ray sources from \citet{Nagao2006}.  These objects are located at systematically lower \civ/\heii\ ratios compared with the emission line selected Type~II QSO sample.  
Similarly, Type~II AGN at redshift 1.5$<$ z $<$3 in \citet{Mignoli2019} objects occupy a similar area of the diagrams.  

We compare the observed line ratios with the predictions from photoionization and shocks models described in Section~\ref{sec:models}.  The emission line ratios are not corrected by dust attenuation, and the impact of this is not expected to be significant \citep{Feltre2016, Hirschmann2019}.  Emission line ratios based on \civ\ and \heii\ are useful diagnostics to distinguish between nuclear activity and star formation \citep[e.g.,][]{Feltre2016}, as the relatively harder AGN ionizing spectrum compared to that of typical star-forming galaxies results in lower \civ/\heii\ ratios.  Figure~\ref{figlineratio} shows that the Type~II AGN models have very distinct \civ/\heii\ ratios, with Log(\civ/\heii)$<$0.5.  
In the model of \citet{Mignoli2019}, a smaller inner radius and microturbulence are adopted.  These adjustments increase the number of continuum photons that can be absorbed by the gas, enhancing the resonant transitions and therefore increasing the intensity of the high ionization lines such as \civ\ and \nv\
\citep[See also][]{Kraemer2007}.
These models typically reproduce well the line ratios in narrow line AGNs \citep[e.g., ][]{Nagao2006,Mignoli2019}.  
Figure~\ref{figlineratio} shows, however, that the line ratios for S82-20 and the A13 Type~II QSO sample are systematically higher than even the most extreme AGN models, but not high enough to be reproduced by models with photoionization by only stars.  
For both S82-20 and the A13 Type~II QSO sample, the \nv/\civ\ and \civ/\ciii] ratios, on the other hand, are consistent with photoionization by either AGN or stars.  For ionization by stars, the \civ/\ciii] requires a relatively large metallicity of the gas, with $Z\gtrsim 0.017$.  For the AGN models, the metallicity constraints are mostly driven by the \nv/\civ\ ratio, which requires $Z\sim 0.008-0.017$.  

The left panel of Figure~\ref{figlineratio} shows that S82-20 and the A13 Type~II QSOs have high values of the \ovi/\civ\ ratios (larger than 0.2), not consistent with photoionization by stars.  These high values can only be explained by photoionization from an accretion disk.  This is because O$^{5+}$ ions require photons with energy at least 113.9~eV, and the spectra of typical massive stars are not hard enough to produce significant amounts of these high energy photons.  We note that line ratios involving the \ovi\ emission lines have to be regarded as lower limits, as at $z\gtrsim 2$, the \ovi\ line fluxes can be affected by absorption in the \lya forest.  Correcting for this absorption would move the points to the right, further away from the SF region.  As Figure~\ref{figlineratio} shows, the metallicity in the AGN models is not constrained by this ratio.

The observed high emission line ratios (Log(\civ/\heii) $\geq$ 0.5) could be produced in AGN NLRs with extreme conditions: very high ionization parameters and very low metallicity. Such models, however, would still not able to reproduce the \civ/\heii\ vs \civ/\nv\ ratios of both S82-20 and the A13 sample, as shown in Figure~\ref{figlineratio}. The AGN models shown in Figure~\ref{figlineratio} assume  a power law index $\alpha=-1.7$. A harder ionization spectrum ($\alpha=-2.0$) can produce even higher \civ/\heii\ ratio, but still would not the \civ/\heii\ vs \civ/\nv\ ratios. See \citet{Mignoli2019} for the full parameters of the models.
In the following sections, we consider various possibilities that could explain the observed properties of S82-20.

\subsection{Type~I+Type~II spectrum}
The high values of the \civ/\heii\ line ratios observed for the A13 sample and S82-20 are similar to those typically measured in the broad line regions of QSOs \citep{Berk2001,Alexandroff2013,Nagao2006_b}.
\citet{Alexandroff2013} suggest that a possible explanation for these large values is that the measurement of the narrow--component fluxes are contaminated by some partially-unobscured emission from the broad component.  This explanation, however, does not seem to apply to S82-20.  In order to increase the \civ/\heii\ line ratio to the observed values, at least 55\% of the \civ\ emission in the $\pm$900 km s$^{-1}$ range would have to come from a broad component.  If this were the case, however, we would have expected to detect the broad component in, e.g., the much brighter \lya emission line.  Additionally, although this explanation could be viable for the A13 objects in which the obscuration is modest (see discussion above), in Figure~\ref{figLyaCIV} we show that the continuum of S82-20 is substantially fainter than in the A13 objects, for the given \lya luminosity, implying that it is substantially more absorbed.

Another possibility is that we observed S82-20 in a temporary, low state.  If the accretion rate of SMBH had decreased recently, the UV continuum, the broad emission line component, and the re-emitted infrared will decrease \citep[e.g.][]{ MacLeod2016, Yang2018}.
The emission originating from the lower density NLR may have a delay response due to its longer recombination timescale \citep{Denney2014, LaMassa2015}.  
Most reported changing-look AGNs do show the Type~I spectral features at some point, since the variation of the broad emission component is a decisive element.  If such changing of the central engine occurs in a Type~II AGN, the accretion disc will be already blocked from the line of sight.  Therefore we may not observe variations in the UV and optical spectrum.  The first observation (Palomar; October 2013) and the most recent (LBT; October 2020) are about 1.7 years apart in the object's rest frame.  The two spectra show similar EWs, line shapes, and line ratios.  If the emission lines are emitted from the NLR, the high \civ/\heii\ ratio requires high ionization parameter and microturbulence as described in Section~\ref{sec:models}.  It is unclear, however, whether a lower state AGN would have enough energy to constantly drive this microturbulence.  Therefore, with the available data, we can not determine whether S82-20 is a changing look AGN.  

\subsection{Type~II+Stars spectrum}
Another possibility is that we are observing the combined emission from gas ionized by both star-formation and AGN. 
In this scenario,  the \ovi\ and \nv\ emission lines are excited by the AGN spectrum, while both AGN and young stars can produce the \civ\ emission.
In addition to reproducing the unusual line ratios, this explanation would also help to explain the faintness of S82-20 in the IR, as in this case the luminosity of the narrow \lya emission would not be a good proxy for the intrinsic ionizing power of the accretion disk.  In this scenario, the observed blue continuum would be the result of SF, and the AGN-powered line luminosity would be decreased.

The \civ\ and \ciii] EWs, in addition to the line ratios, can be used to identify the composite origin of the emission lines, as done by e.g., \citet{Nakajima2018} and \citet{Hirschmann2019}.  We do this in Figure~\ref{figEWCHe}, where we show the classification regions introduced by the two groups.  These diagnostics are not conclusive.  According to the \civ-based plot (left panel in Figure~\ref{figEWCHe}), S82-20 as well as the A13 objects, fall in the ``composite" part of the diagram.  The \ciii]-based diagnostic, however, shows that the A13 objects are located across the separation between AGN and composite spectra, while S82-20 is consistent with AGN photoionization.  This discussion highlights how the availability of multiple line ratios often reveals a situation that is more complex than what would appear with limited information.
In a composite spectrum scenario, at least $\approx 50$\% of the \civ\ luminosity would need to be the result of photoionization by stars, in order to move the line ratios from the most extreme AGNs to the observed position .  

We can attempt to place a constraint on the stellar mass of the host galaxy, using the two-component SED fitting procedure by \citet{Bongiorno2012}, in which the observed optical-to-NIR SED is fitted with a large grid of models created from a combination of AGN and host galaxy templates.   
In particular, for the AGN component, this procedure adopts the \citet{Richards2006} mean QSO SED, extinguished by applying a SMC-like dust-reddening law \citep{Prevot1984} of the form:$A_{\lambda}/E(B-V) = 1.39\times \lambda_{\mu {\rm m}}^{-1.2}$, while for the galaxy component, a library of synthetic spectra generated using the \citet{Bruzual2003} stellar population synthesis models with declining star formation histories (SFR $\propto e^{-t_{age}/\tau}$) with e-folding times, $\tau$, ranging from 0.1 to 30 Gyr and a grid of ages ranging from 50 Myr to 9 Gyr.  Dust extinction of the galaxy component is taken into account using the Calzetti law \citep{Calzetti2000}.  For more details, we refer the reader to \citet{Bongiorno2012}.The results of the SED fitting code suggest that the host galaxy is a relatively massive, actively star-forming system, with $\log(M_*/M_{\odot})=10.6^{+0.2}_{-1.4}$ and a star-formation rate $\log(SFR)=1.5^{+0.5}_{-0.1}$.  The AGN continuum results completely obscured in the rest-frame UV, with an estimated E(B-V)=9.  

An interesting question is where the obscuration of the AGN continuum takes place, as it can happen either close to the disk, in the nuclear torus, or in the host galaxy, following a dusty merger phase \citep[as suggested by, e.g.,][]{Hopkins2008}.  The available observations disfavor the latter interpretation.  First, the UV continuum is very blue, with $\beta = -3.3^{+0.6}_{-0.9}$.  Although in principle such blue values of $\beta$ can be produced by young, low-metallicity stars (e.g.  \citet{Schaerer2002, Bouwens2010}), this color would imply that the stellar continuum is not attenuated.  Second, the bright \lya luminosity is difficult to reconcile with a dust-rich environment.  If the observed attenuation were due to dust distributed on galaxy-scale, we would expect the \lya to be substantially suppressed.  Consequently, it appears more likely that the obscuration of the AGN continuum and the broad line region happens close to the accretion disk, in a galaxy with an overall low dust content that allows the escape of a substantial amount of \lya.

\subsection{Shocks}
It is worth exploring an additional mechanism that can result in strong high-ionization emission lines: high velocity radiative shocks.  Shocks could result from gas accelerated either by the explosion of massive stars associated with regions of active star-formation \citep[e.g., ][]{Yadav2017, Rosdahl2017}, or in the high-velocity outflows often associated with AGN.  In Figure~\ref{figlineratio} we show the line ratios predicted by the shock models with purple lines.  Models with higher metallicity ($\sim$0.5 - 1 Z$_\odot$) and a velocity of 200~km~s$^{-1}$ are broadly consistent with most of the A13 Type~II QSO candidates as well as with S82-20, in all diagnostic plots considered in Figure~\ref{figlineratio}.  Therefore even if the mixing source scenarios reproduce the emission line ratio, an unknown fraction of flux can be still contributed from shocks.

Although the shock explanation well reproduces all the emission line ratios, it is not clear whether the entire luminosity of the emission lines can be explained with shocked gas.  Additionally, this interpretation still does not shed light on the source of energy that powers the required high velocity gas, as outflows of velocities up to 1000s km s$^{-1}$ are observed in both AGN and star-forming galaxies \citep{Diamond-Stanic2012, Harrison2014, Zakamska2014}.

\section{Conclusions}\label{sec:conclusion}
We presented a luminous $z=3.08$ broad-band selected \lya\ emitting object, S82-20, identified through the excess flux in the $g$-band compared to adjacent filters using deep SDSS Stripe 82 data.  In this paper we describe the spectroscopic data obtained to confirm the redshift of the source, as well as the spectral energy distribution derived from archival near and mid-infrared data.  We use this information to place constraints on the nature of the source ionizing the gas in this object.

The rest-frame UV spectrum shows highly ionized emission lines, including the \civ $\lambda$1550 doublet, \heii $\lambda$1640 and \ovi $\lambda$1032 doublet.  The high \lya luminosity ($3.5\times 10^{44}$ erg s$^{-1}$), the absence of broad wings associated with the emission lines, the large emission-line EWs, and the detection of the \ovi\ doublet strongly suggest the interpretation that S82-20 is an obscured Type~II QSO.

However, some evidence suggests that the Type~II QSO interpretation may not be sufficient to explain all the observables we presented in the paper.  The \civ/\heii\ vs.  \nv/\civ\ ratio is not fully reproduced by AGN photoionization models alone.
Additionally, S82-20 is not detected at wavelengths longer than 2$\mu$m, in tension with the expectation that most of the accretion disk luminosity should be absorbed by the dusty torus and re-emitted isotropically in the far-IR.
To fully explain these features, S82-20 may require either some contribution from metal rich star-formation or high velocity shocks.  The lack of broad lines could be explained if S82-20 is a QSO in a temporary low state. This scenario, however, may still have problems in reproducing the ratio of the narrow \civ/\heii\ lines. 

Whether the explanation of the observed emission line-ratios in S82-20 is photoionization by a combination of AGN and SF or shock heated gas, this object is a rare example of the short phase of the life of a massive galaxy, in which active SF and accretion onto a super-massive BH coexist \citep{Hopkins2008}.  This phase is supposed to be characterized by large obscuration of both the SF episode and the AGN.  Our result, however, may have identified a new phase, in which only the nuclear region is heavily obscured, while the galaxy lacks a diffuse dust component.  

We conclude by noting that this object was selected only because the \lya emission line made it bright enough to be selected in the $g-$band.  Had it been at a different redshift, we may not have selected it.  Additionally, this object would have been clearly missed in rest-frame continuum surveys of QSOs, as Figure~\ref{figsed} shows that it is approximately an order of magnitude fainter than known Type~II QSOs at similar redshifts.

\section*{Acknowledgements}
This research has made use of the NASA/IPAC Infrared Science Archive, which is funded by the National Aeronautics and Space Administration and operated by the California Institute of Technology.
Some of the observations reported in this paper were obtained with the Southern African Large Telescope (SALT) under program 2018-1-MLT-011 (Matthew Hayes).
This paper used data obtained with the MODS spectrographs built with funding from NSF grant AST-9987045 and the NSF Telescope System Instrumentation Program (TSIP), with additional funds from the Ohio Board of Regents and the Ohio State University Office of Research.
M.H.  is fellow of the Knut and Alice Wallenberg Foundation.
AF acknowledges the support from grant PRIN MIUR2017-20173ML3WW$\_$001.
MM and PV acknowledge support from the National Research Foundation of South Africa.

\section*{Data Availability}
The data used in this article will be shared on reasonable request to the corresponding author.



\bibliographystyle{mnras}
\bibliography{S8220}







\bsp	
\label{lastpage}
\end{document}